\documentclass[10pt]{iopart}
\usepackage{graphicx}

\begin{document}

\title[Jet tomography]{Jet tomography}

\author{Ivan Vitev\dag }

\address{\dag\ Iowa State University, Department of Physics and Astronomy, \\
12 Physics Building A330, Ames, IA 50011, USA }

\ead{ivitev@iastate.edu}

\begin{abstract}
I summarize  the recent advances in jet tomographic studies 
of cold and hot nuclear matter based on perturbative QCD calculations 
of medium-induced  gluon bremsstrahlung. Quantitative 
applications to ultrarelativistic heavy ion reactions at RHIC indicate
the creation of a deconfined state of QCD with initial energy density 
on the order of 100 times cold nuclear matter density. 
\end{abstract}


{\bf Introduction.}
Tomography is the study of the properties of matter through the 
attenuation pattern of a calibrated flux of fast particles that  
lose energy via multiple elastic and inelastic scatterings.  
For ultrarelativistic heavy ion reactions this technique 
was first discussed for vector mesons but has only recently been 
generalized to hard partonic processes which dominate the 
experimentally accessible moderate- and high-$p_T$ region of 
hadron production~\cite{Lovas:ge}. Jet tomography  
extracts the nuclear matter density from the  
modification to the differential hadron multiplicities
$R_{AB}(b) = \Big( \frac{dN^h_{AB}(b)}{dy d^2p_T}\Big) {\big/} 
\Big( \frac{T_{AB}(b) d\sigma^h_{pp}}{dy d^2p_T} \Big)$. The overlap
function $T_{AB}(b)$ reflects the reaction geometry and the elementary 
nucleon-nucleon cross sections are calculable
to leading and next-to-leading order~\cite{Eskola:2002kv} based 
on the successful perturbative QCD factorization 
approach~\cite{Bodwin:1984hc}.  The potentially dramatic observable 
effects of jet quenching~\cite{Wang:xy},  
$R_{AB} \sim \exp \big( -\int_0^L \, dz^\prime \sigma_{\rm abs}
(z^\prime) \rho(z^\prime) \big)$,  in $A+A$ 
versus $p+p$ collisions even for dynamically expanding systems
come from the nuclear size enhancement $\propto r_0 A^{1/3}$ 
and the large initial energy and parton number densities of the
quark-gluon plasma (QGP) created in the interaction region. In addition 
to the suppression of the single inclusive spectra~\cite{Adler:2003qi} 
and the disappearance of the back-to-back di-hadron 
correlations~\cite{Adler:2002tq} in central and semi-central $Au+Au$
reactions, parton energy loss plays a major role in the generation of 
the high-$p_T$ azimuthal asymmetry~\cite{Wang:2000fq}  and the 
enhancement of the moderate-$p_T$ baryon to meson ratios~\cite{Vitev:2001zn}. 
The final state origin of many of these effects is now confirmed 
via the control $d+Au$ measurement and its comparison to theoretical 
predictions~\cite{Arsene:2003yk,Zhang:2001ce}.

The first calculations of the Landau-Pomeranchuk-Migdal 
(LPM) destructive interference effect in QCD~\cite{Wang:1994fx} 
already indicated that  medium induced non-Abelian bremsstrahlung 
dominates over the elastic energy loss.
The current techniques for the computation of the gluon radiative 
spectrum  $ \frac { \omega dN^g }{ d\omega } $ and the 
mean energy loss  $\Delta E$  have been reviewed in~\cite{Gyulassy:2003mc}  
and include  a path integral approach~\cite{Zakharov:1996fv}, 
an effective 2D Schr\"oedinger equation~\cite{Baier:1996kr}, exact Reaction 
Operator formalism~\cite{Gyulassy:2000er} and a twist expansion 
series~\cite{Zhang:2003yn}. 
For static non-Abelian media the mean energy 
loss depends quadratically on the nuclear size, $\Delta E = -
\frac{C_R \alpha_s}{4}\frac{\mu^2}{\lambda_g} L^2  \ln \frac{2E}{\mu^2 L}
+ \cdots \;$,  and measures the gluon transport coefficient 
$\hat{q} = \frac{\mu^2}{\lambda_g}$. In dynamically  
expanding plasmas the  radiative spectrum  and the mean energy loss 
can be related to the soft gluon rapidity density, 
$\Delta E = - \frac{9  C_R \pi \alpha_s^3}{4} 
\frac{1}{A_\perp}  \frac{dN^{g}}{dy}  L  \, 
\ln \frac{2E}{\mu^2  L } + \cdots \;$.
The outlined path length dependence holds for both small and  
large numbers of scatterings but $\frac{\omega dN^g}{d\omega}$  
is more sensitive to the few semihard kicks that hard partons
undergo as they traverse the medium~\cite{Gyulassy:2000er}.
Major corrections to the above analytic  approximations arise 
from the finite kinematics 
which is particularly important at SPS and RHIC and even 
at the LHC for  $p_T < 50$~GeV. These can be numerically  
evaluated as in~\cite{Gyulassy:2000er}.

\begin{figure}[t!]
\begin{center}
\includegraphics[width=2.in,height=1.5in]{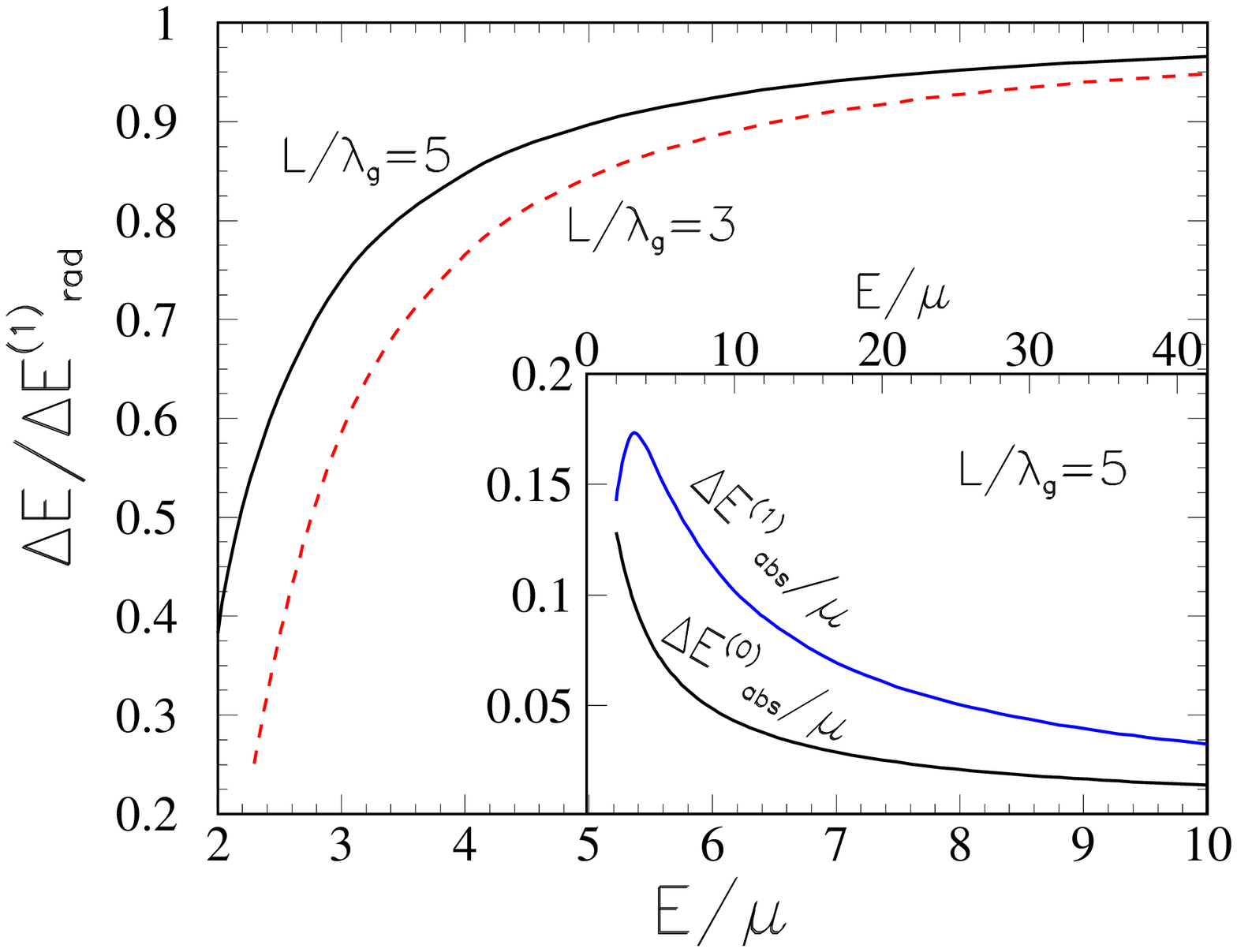}
\hspace*{0.9cm}\includegraphics[width=2.3in,height=1.6in]{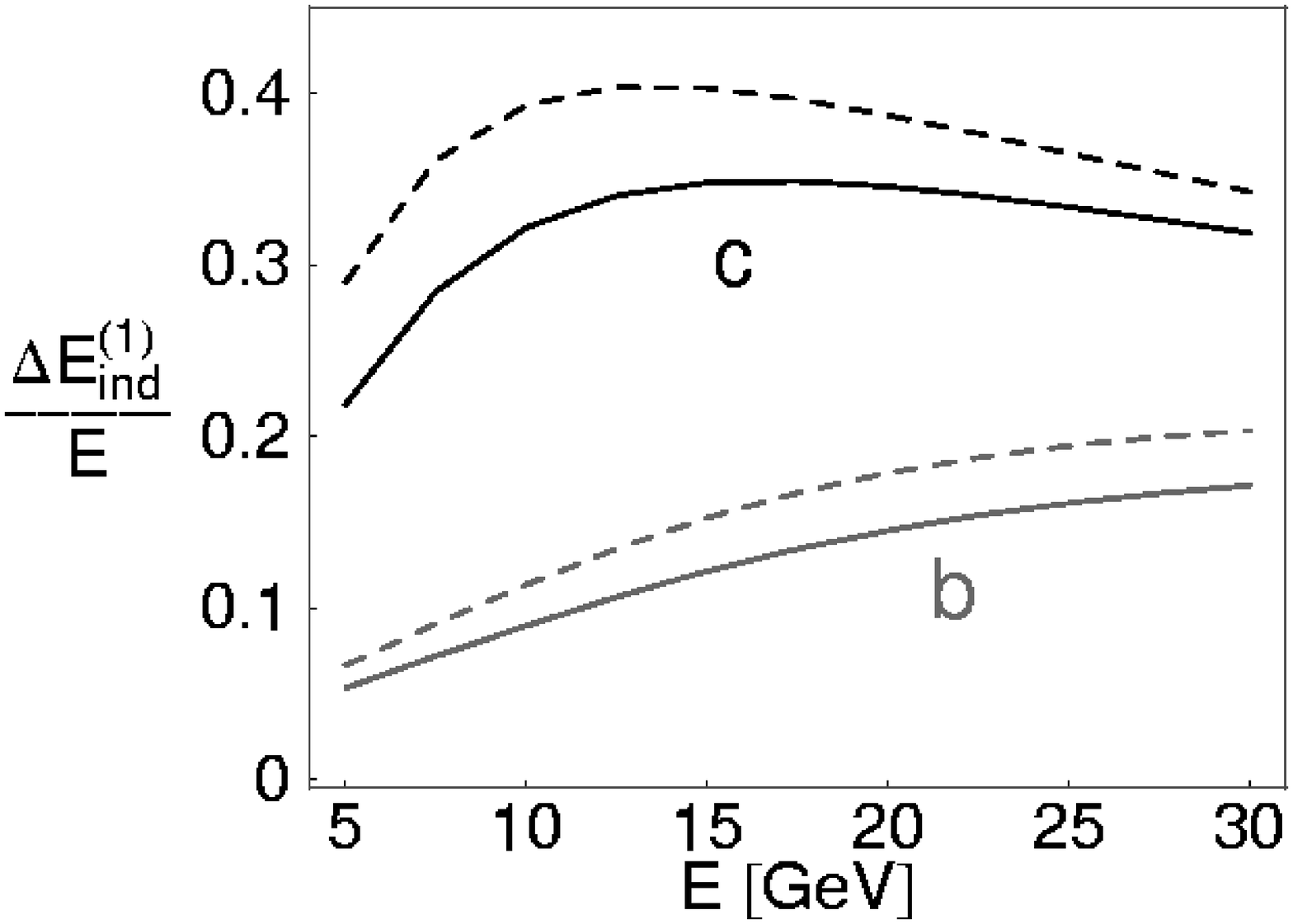} 
\end{center}
\vspace*{-0.3cm}
\caption{(a) From~\cite{Wang:2001cs}: 
reduction of the medium induced radiative energy 
loss via gluon absorption as a function  of $E / \mu$. 
(b) From~\cite{Djordjevic:2003zk}: jet energy dependence of 
the mean  fractional energy 
loss $\Delta E/E$. The sensitivity on the plasmon mass $\omega_{\rm pl}$ 
and the heavy quark mass $M_q$ is illustrated.}
\label{fig1-2}
\end{figure}

\begin{figure}[b!]
\begin{center}
\vspace*{-.3cm}
\includegraphics[width=2.in,height=1.5in]{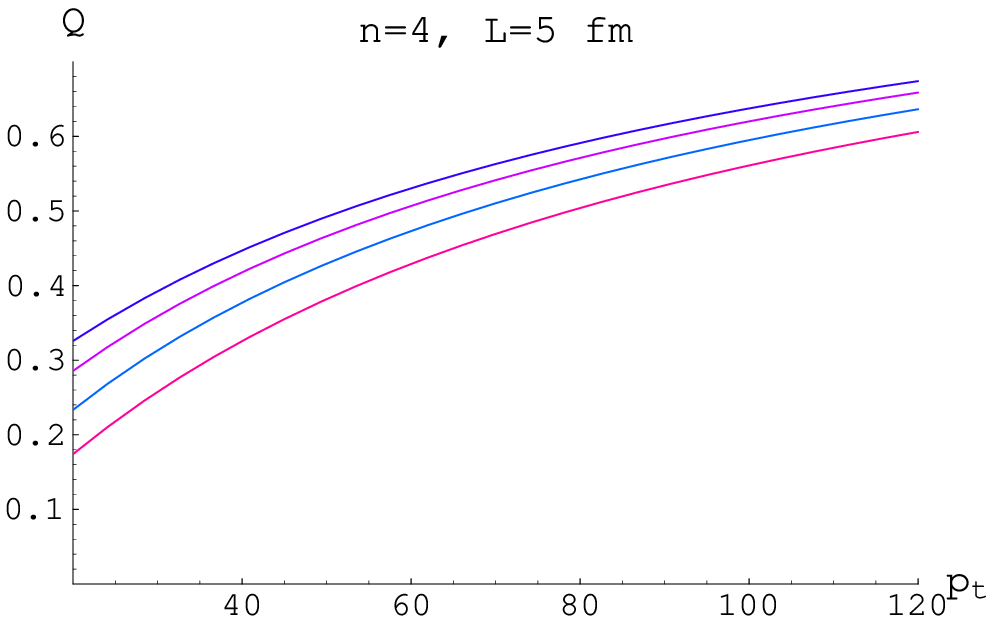}
\hspace*{0.9cm}\includegraphics[width=2.1in,height=1.9in]{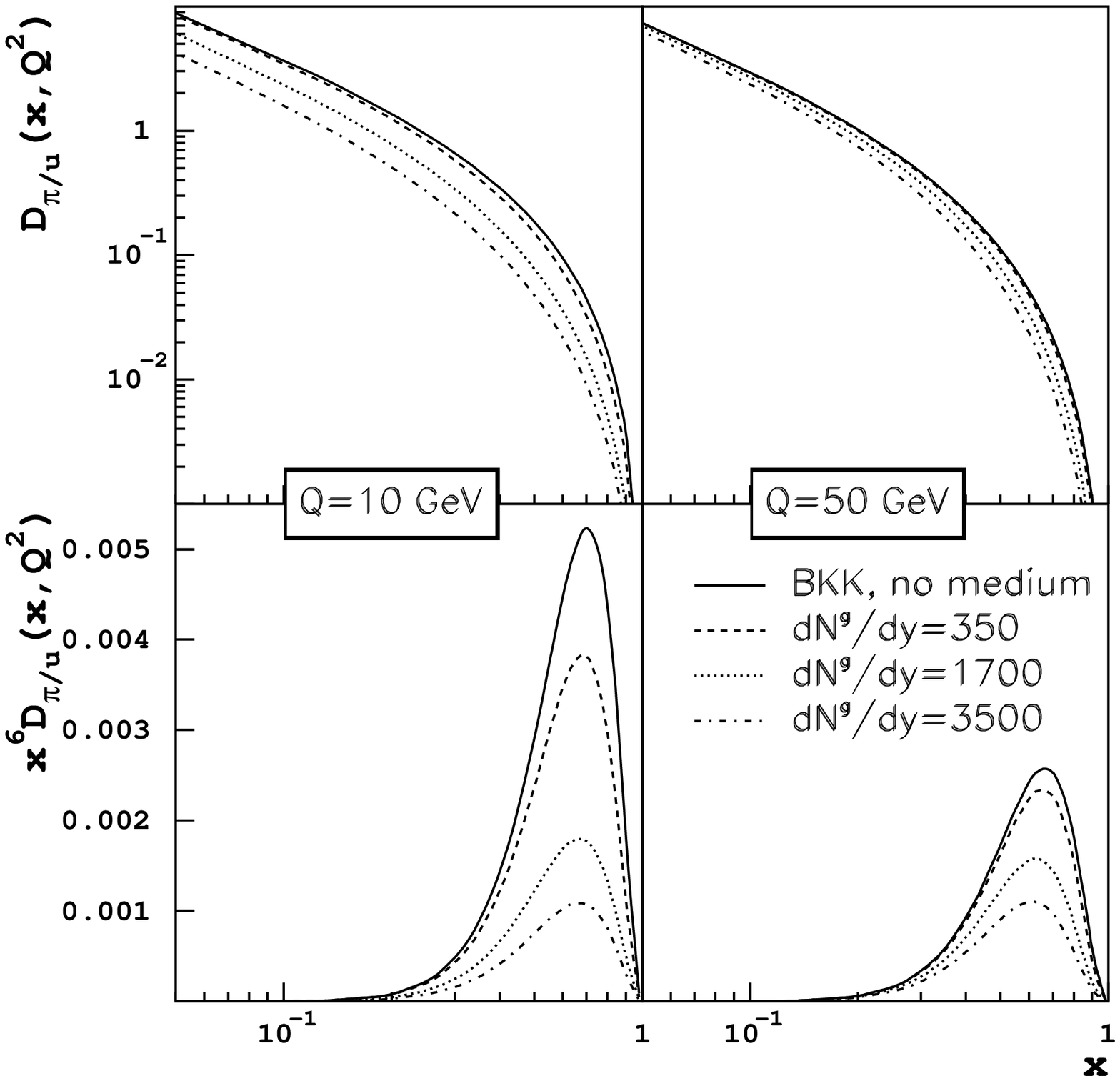} 
\end{center}
\vspace*{-0.3cm}
\caption{ (a) From~\cite{Baier:2001yt}: typical 
asymptotic $p_T$ dependence of the 
partonic quenching factor  for LHC energies. 
(b) From~\cite{Gyulassy:2001nm}: effective kinematic modification 
of the fragmentation  functions versus $x = p_h/p_c$.}
\label{fig3-4}
\end{figure}

For QGP in a local thermal equilibrium all 
dimensional scales can be related to the 
temperature: $ \omega_{\rm pl}^2 \sim \mu^2 \propto T^2$, 
$\rho_g \propto T^3$ and  $\epsilon \propto T^4$. Moreover,  
gluon absorption~\cite{Wang:2001cs} can significantly modify 
the radiative spectrum and the mean energy loss
for small jet energies as shown in Fig.~\ref{fig1-2}a. 
The average energy gain from gluon absorption to first order 
in opacity,  $\Delta E_{\rm abs} = 
\frac{C_R  \pi \alpha_s}{3}\frac{ T^2}{\lambda_g E} L 
\ln \frac{\mu^2 L}{T} + \cdots \;$,  is power suppressed 
for large $E$ but will manifest itself for $E < 5 \mu$ 
via weaker hadron quenching.  The reduction 
of available phase space for bremsstrahlung off heavy quarks, the dead cone 
effect~\cite{Dokshitzer:2001zm}, and the plasmon mass $\omega_{\rm pl}$  
Ter-Mikayelian effect have been systematically incorporated 
in the medium induced energy loss formalism~\cite{Djordjevic:2003zk} 
and shown to shorten the  gluon formation times, 
$ \tau_f^{-1} \rightarrow  \tau_f^{-1} + 
\frac{\omega_{\rm pl}^2 + \omega^2 M_q^2 /E^2 }
{2 \omega}$,  and regulate the color current propagators, 
$\frac{k_\perp}{k^2_\perp} \rightarrow \frac{k_\perp}{k^2_\perp 
+ \omega_{\rm pl}^2 + \omega^2 M_q^2 /E^2} $. 
The dependence of $\frac{\Delta E}{E}$ on the heavy quark mass is
illustrated in  Fig.~\ref{fig1-2}b. In addition to a 
reduction of the mean energy loss the path length dependence
of $\Delta E$ in static plasmas was found to be closer to 
the linear incoherent 
Bethe-Heitler regime~\cite{Djordjevic:2003zk,Zhang:2003wk}.

Approximate ways of estimating the observable jet quenching effect 
on final state hadrons include a suppression of the hard partonic 
cross section~\cite{Baier:2001yt} or an effective 
attenuation of the fragmentation  
functions~\cite{Gyulassy:2001nm}, 
$D_{\rm eff}(z) = \int d\epsilon \, P(\epsilon) \frac{1}{1-\epsilon} 
D\big(\frac{1}{1-\epsilon} \big)$. Here 
$P(\epsilon)$ is the probability of fractional energy loss 
$\epsilon = \sum_i \frac{\omega_i}{E}$
due to multiple gluon emission~\cite{Baier:2001yt,Gyulassy:2001nm}.
These incorporate the same kinematic modifications to the 
leading hard parton which originate from the medium induced  
gluon bremsstrahlung at a timescale  $ t_0 = \frac{1}{Q}  
< t_{\rm rad} <  
t_{\rm had} =  
\frac{2 z (1-z) E_{\rm jet}}{\Lambda_{\rm QCD}^2}  $.    
The former approach is illustrated in Fig.~\ref{fig3-4}a for  
a static QGP with $\hat{q} = 
\frac{\mu^2}{\lambda_g} =  1\; \frac{\rm GeV^2}{\rm fm}$ and the latter 
is exemplified in Fig.~\ref{fig3-4}b for $D_{u/\pi}(z,Q^2)$. 
Quantitative applications, however, require the incorporation of 
energy loss calculations within the full perturbative QCD hadron 
production formalism~\cite{Vitev:2002pf}.

\begin{figure}[!b]
\begin{center}
\includegraphics[width=2.in,height=1.5in]{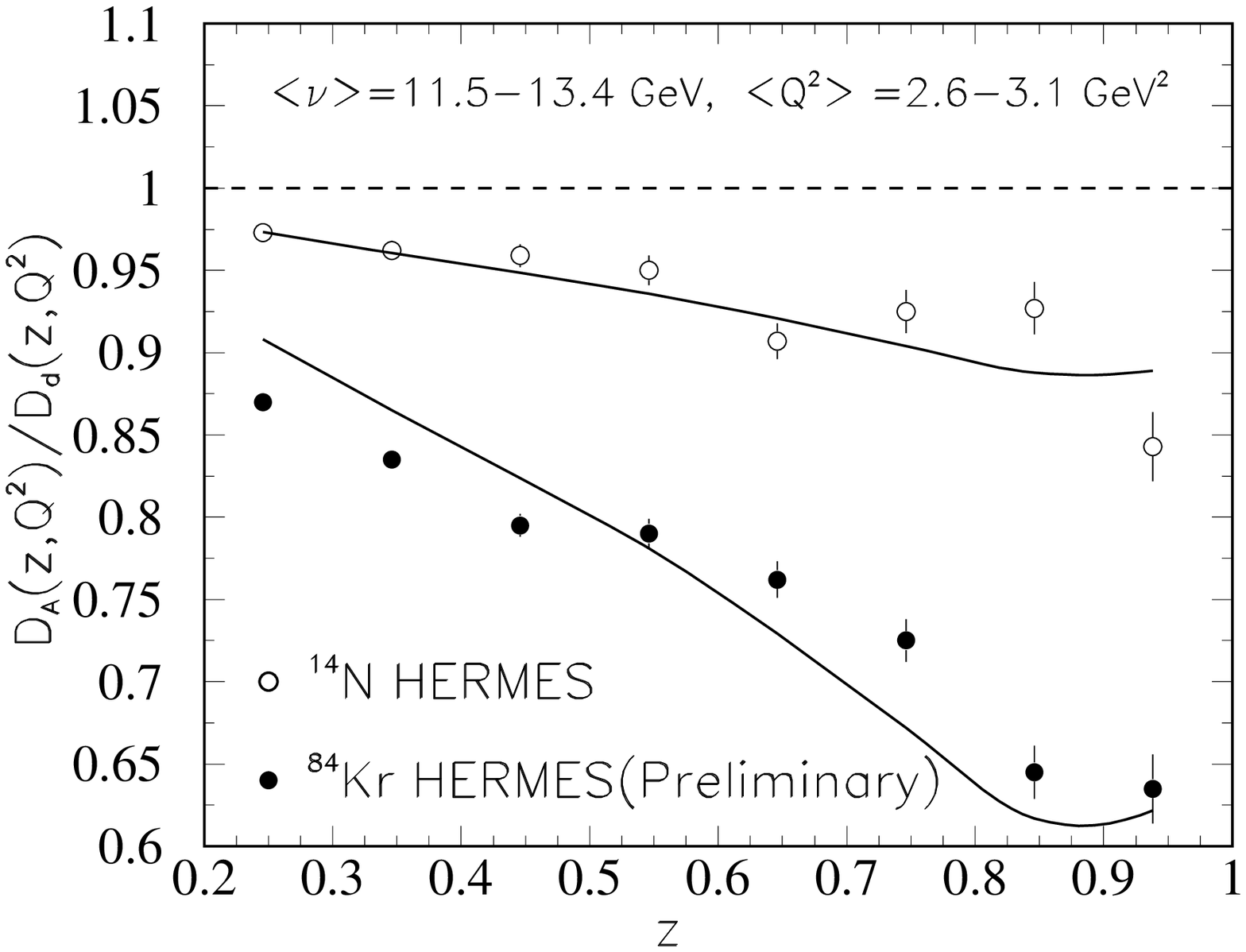}
\hspace*{0.9cm}\includegraphics[width=2.in,height=2.1in]{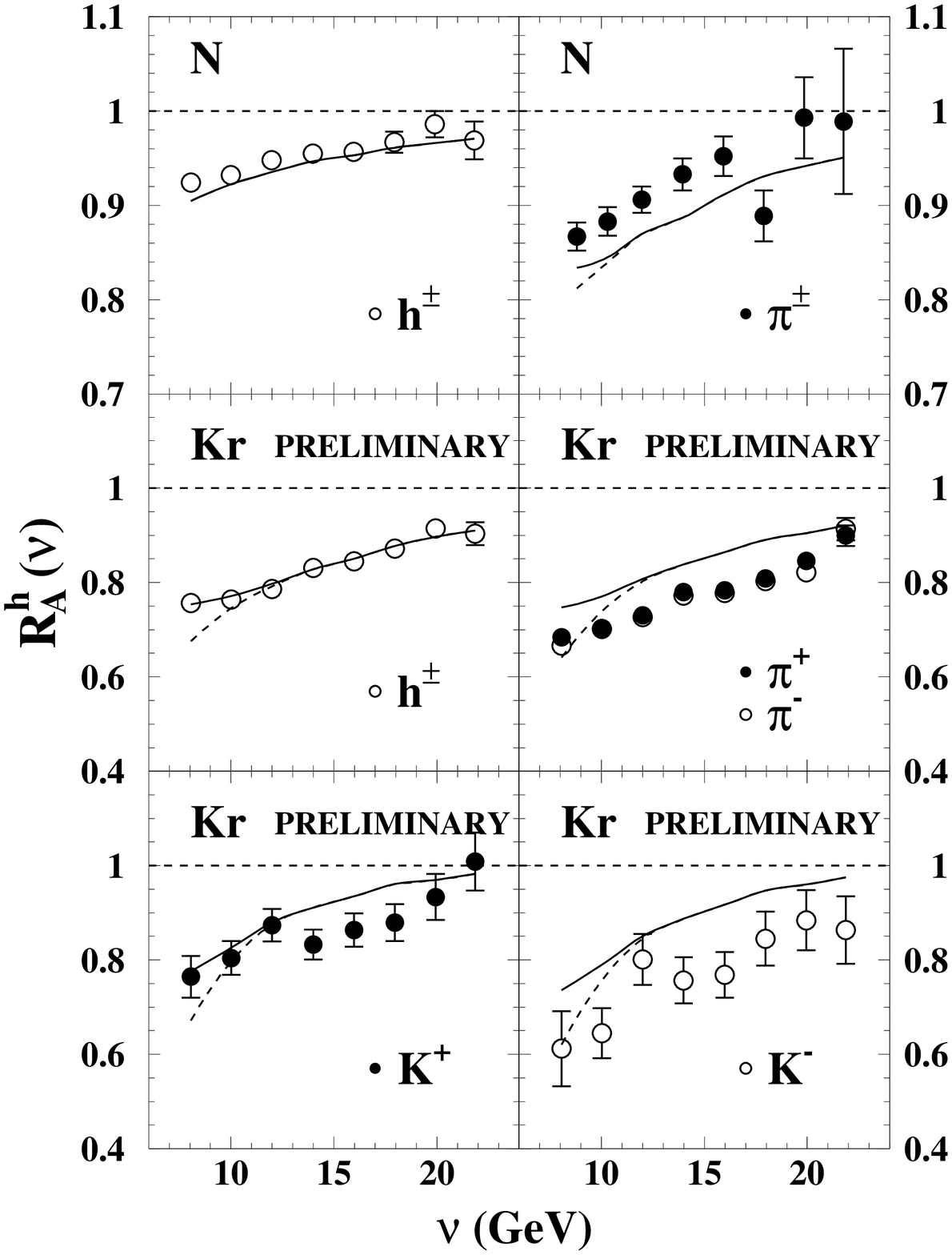} 
\end{center}
\vspace*{-0.3cm}
\caption{ (a) From~\cite{Wang:2002ri}: 
evidence for the quadratic dependence of 
$\Delta E$ on the $L \propto A^{1/3}$ system size in semi-inclusive 
DIS. Hadron quenching is shown versus  $E_h / \nu$. 
(b) From~\cite{Arleo:2003jz}: calculated $\nu = E-E^\prime$ 
dependence of  nuclear  suppression with 
$\hat{q}=\frac{\mu^2}{\lambda_g}$ fixed from Drell-Yan data.}
\label{fig5-6}
\end{figure}


{\bf Applications of jet tomography.}
Semi-inclusive deeply inelastic scattering (DIS) is an ideal 
probe of final state quark energy loss in cold nuclear matter. 
The known lepton $E_l,E_l^\prime$ and hadron 
$E_h$ energies allow for an approximate reconstruction of 
the effective modification to the measured cross section
$ \sim {D^{eA}_{h/q}(z)}/{D^{ep}_{h/q}(z)}$, where 
$z = E_h/\nu$. Induced gluon bremsstrahlung  was shown to result in an 
effective shift,  $ \Delta z \propto \alpha_s A^{2/3} \ln \frac{1}{x_B}$,   
of the fragmentation momentum fraction~\cite{Wang:2002ri}.  
Comparison of HERMES data on the nuclear modification for 
$^{14}N$ and $^{84}Kr$  targets to theoretical calculations 
in  Fig.~\ref{fig5-6}b  provides strong evidence in support of 
the $L^2 $ dependence of $\Delta E$ for static systems.
The absorptive power of the medium for quark jets 
was found to be $ -\frac{dE_q}{dz} =  0.5 \; \frac{\rm GeV}{\rm fm}$.
Constraints on the cold nuclear matter 
transport coefficient can also be obtained from the analysis
of Drell-Yan data on initial state broadening and 
energy loss~\cite{Arleo:2003jz}. The extracted 
$\frac{\mu^2}{\lambda_g} = 0.14 \pm 0.1
\; \frac{\rm GeV^2}{\rm fm}$ can be used to predict 
the final state hadronic suppression.  Figure~\ref{fig5-6}b 
shows the consistency of this approach  on the example
of the $\nu$-dependent quenching.

\begin{figure}[!t]
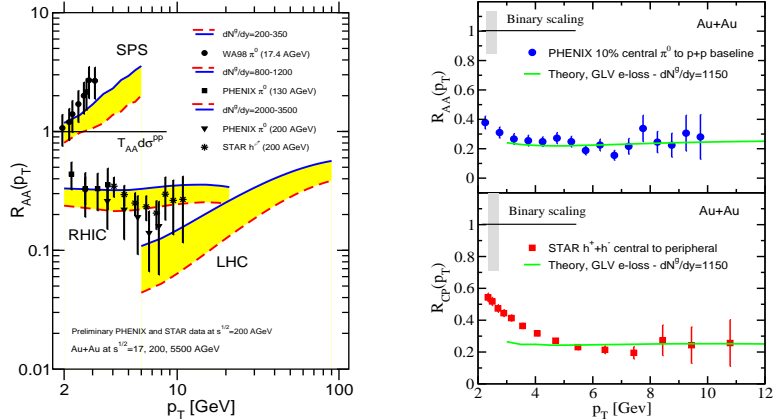

\begin{center}
\includegraphics[width=1.8in,height=2.2in]{VitevPred.eps}
\hspace*{0.9cm}\includegraphics[width=1.8in,height=2.2in]{VitevCompare.eps} 
\end{center}
\vspace*{-0.3cm}
\caption{ (a) From~\cite{Vitev:2002pf}: predicted 
$\sqrt{s_{NN}}$ and $p_T$ dependence of jet quenching  for finite 
kinematics and  finite system size. 
(b) From~\cite{Vitev:2002pf}: comparison of the 
calculated approximately constant $R_{AA}$
for $\pi^0$ and $h^\pm$ to finalized experimental data.}
\label{fig7-8}
\end{figure}

The initial discovery of a qualitatively large 
high-$p_T$ particle deficit at RHIC 
$\sqrt{s_{NN}} = 130$~GeV~\cite{Adcox:2001jp} has revived the interest in 
hard physics in relativistic heavy ion reactions. 
However, it is the recent advances in the perturbative QCD 
description of multi-parton dynamics that have taken
jet tomography to a new and quantitative 
level~\cite{Gyulassy:2003mc}.
Figure~\ref{fig7-8}a illustrates the  predicted
$p_T$ dependence of $R_{AA}$ versus $\sqrt{s_{NN}}$ for central 
$Au+Au$ collisions~\cite{Vitev:2002pf}. The relative importance of 
the  initial state multiple elastic 
scattering~\cite{Zhang:2001ce} relative to the final state radiative 
energy loss was found to decrease from SPS to the LHC. At high 
$Q^2\sim p_T^2$ the  $x < 0.1-0.01$   shadowing  
vanishes~\cite{Qiu:2004qk} but the small effects from the $ x > 0.3 $ EMC  
region are expected to persist. Figure~\ref{fig7-8}b  shows 
a good agreement of 
the calculated quenching for $\pi^0$ and  $h^\pm$  with 
$\frac{dN^g}{dy} = 1150$, comparable to the $y=0$ net hadron rapidity
density, with the precision high-$p_T$ data at RHIC~\cite{Adler:2003qi}. 
Ongoing re-analysis of the SPS data~\cite{d'Enterria} is expected to
establish a small hadronic suppression at  $\sqrt{s_{NN}}=17.4$~GeV
and thus improve the agreement with the perturbative calculation 
in a medium of finite, $\frac{dN^g}{dy} = 400 - 500$, soft parton density.

The centrality and rapidity dependence of jet quenching in $A+A$ 
reactions has also been addressed both analytically and 
numerically~\cite{Barnafoldi:2003pi}. 
For 1+1D  expansion hadronic attenuation scales with the
number of participants as $R_{AA}(N_{\rm part}) = \big[1 - 
(1- (R_{AA}^{\rm cent})^{\frac{1}{n-2}}) (N_{\rm part})^{2/3}
/ (N_{\rm part}^{\rm cent})^{2/3}\big]^{n-2}  $ and is insensitive 
to the effective partonic slope $n$. As a function of rapidity
$R_{AA}(y) \approx R_{AA}(y=0)$, in agreement with 
the BRAHMS results~\cite{Arsene:2003yk}. Corrections to these 
estimates arise from the interplay of soft and hard physics 
and the steepening of particle spectra with rapidity 
$|y-y_{\rm mid }|$.

\begin{figure}[b!]
\begin{center}
\includegraphics[width=2.in,height=1.8in]{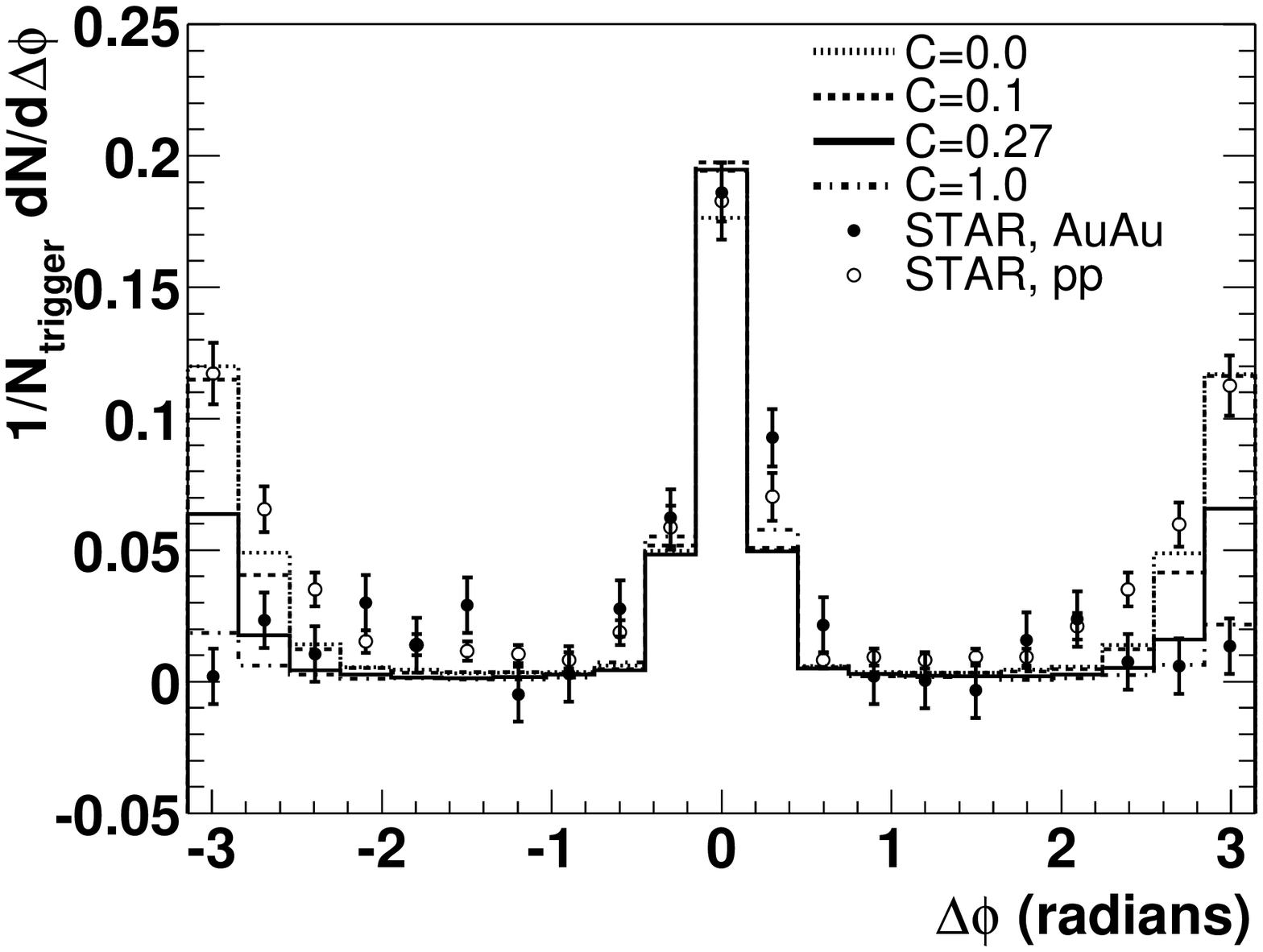}
\hspace*{0.9cm}\includegraphics[width=2.in,height=1.8in]{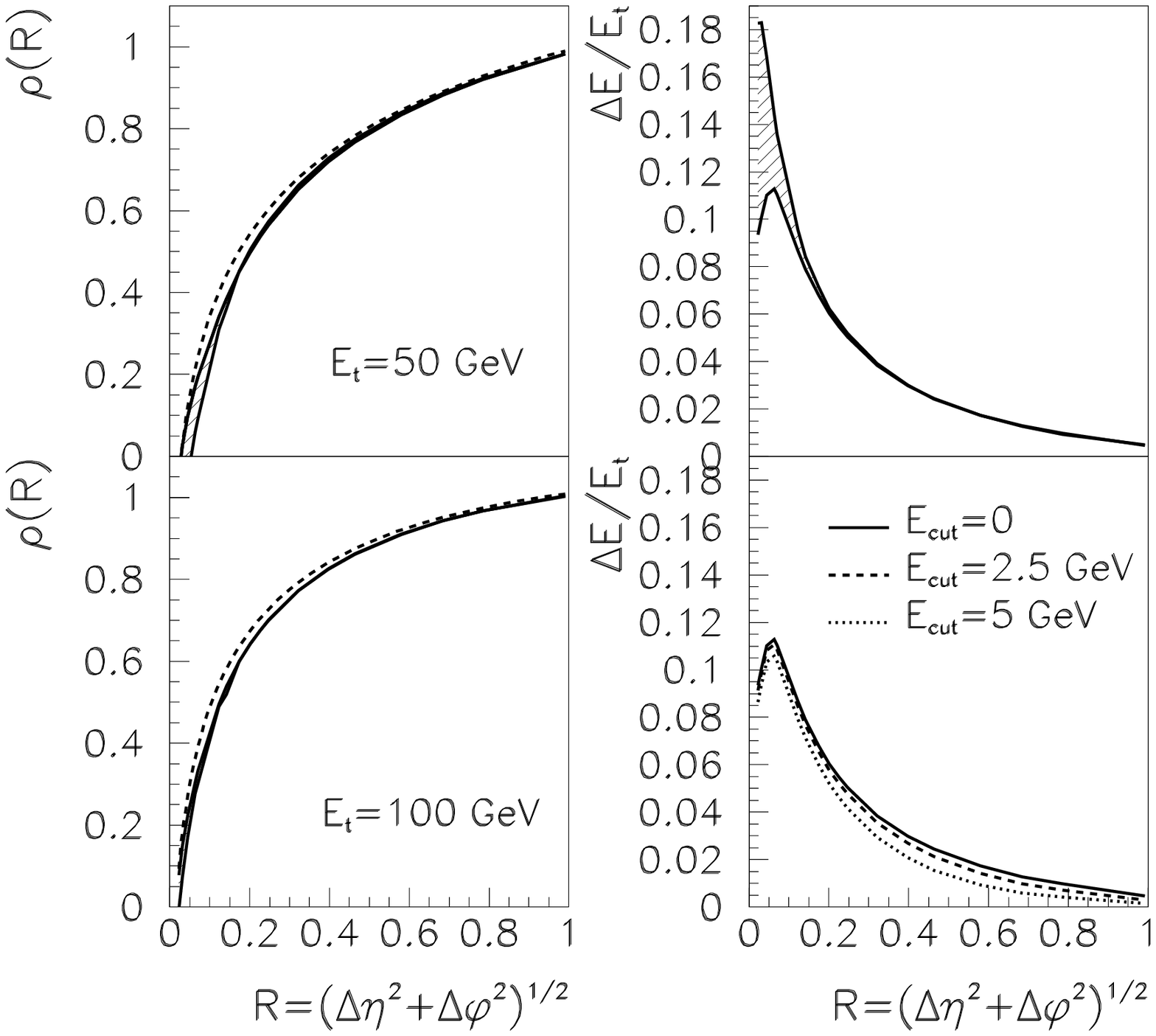} 
\end{center}
\vspace*{-0.3cm}
\caption{ (a) From~\cite{Hirano:2003hq}: disappearance of 
the away side correlation function in a hydro+jet model. 
(b) From~\cite{Salgado:2003rv}: minimal $\sim 5\%$ modification 
to the intra-jet correlations for approximately collinear 
gluon bremsstrahlung.  }
\label{fig9-10}
\end{figure}

Di-hadron 
tomography~\cite{Hirano:2003hq,Qiu:2003pm,Rak:2003ay} is one of 
the new and  promising direction of moderate- and  
high-$p_T$  studies of in-medium jet modification.
The differential cross section 
$\frac{d\sigma^{h_1h_2}}{dy_1dy_2 dp_{T1} dp_{T 2} d \Delta \phi }$ is more 
sensitive to the non-Abelian dynamics when compared to the single
inclusive measurements. Triggering on a large transverse momentum 
particle biases the hard scatter toward the surface of the 
interaction region and the away-side partner  
carries most of the observable effect from the 
interactions in dense nuclear matter.  For the example of the 
correlation function, 
$C_2(\Delta \phi) = \frac{1}{N_{\rm trig}} \frac{dN^{h_1h_2}}{d\Delta \phi}
\approx A_{\rm Near} 
\exp \big(-\frac{\Delta \phi}{2\sigma^2_{\rm Near}} \big) + 
 A_{\rm Far} 
\exp \big(-\frac{(\Delta \phi -\pi)}{2\sigma^2_{\rm Far}} \big)$,
energy loss  reduces the area $\propto A_{\rm Far}$ under the away-side 
peak. Figure~\ref{fig9-10}a  show a simulation of the disappearance 
of the back-to back correlations in a hydro+jet model~\cite{Hirano:2003hq} 
which is qualitatively  consistent with the data. Important 
progress in  di-jet correlation analysis has been made through 
recent studies of the attenuation effect relative to 
the reaction plane~\cite{Rak:2003ay}.  It confirms the $\Delta E \propto L $  
path length dependence of energy loss in an expanding 
quark-gluon plasma~\cite{Wang:2000fq}.  
Di-jet acoplanarity reflects the accumulated mean transverse momentum 
kick $\langle k_T^2 \rangle  = \langle k_T^2 \rangle_{\rm vac} + 
\langle k_T^2 \rangle_{\rm IS} + \langle k_T^2 \rangle_{\rm FS}$  of a 
hard scattered parton pair~\cite{Qiu:2003pm} and is 
measurable~\cite{Rak:2003ay} via the enhanced away-side width 
$\sigma_{\rm Far}$. Preliminary results for $d+Au$ and central 
$Au+Au$  collisions suggest a significant difference in  
the observed effect consistent with  final state interactions in 
media of very different density and scattering power 
$\frac{\mu^2}{\lambda_g}$.
A large $\sigma_{\rm Far}$ enhancement in central $Au+Au$ reactions 
may also indicate a power-law non-Gaussian semi-hard multiple 
scattering~\cite{Gyulassy:2002yv}.

{\bf Future directions in jet quenching studies.}
Improved experimental techniques for jet cone
reconstruction in the high-multiplicity environment of 
relativistic heavy ion reactions~\cite{Rak:2003ay,FWang} 
open new possibilities for differential studies of the 
jet properties. The transverse momentum broadening of 
the interacting parent parton+gluon system is expected to 
modify the intra-jet correlations and widen the jet 
cone~\cite{Gyulassy:2003mc,Salgado:2003rv}. A 
quantitative measure of this effect will be the  
deposition of smaller fractional transverse energy  
$\rho(R) = \frac{1}{N_{\rm jets}} \sum_{\rm jets}
\frac{\Delta E_{T}(R)}{\Delta E_{T}(R=1)}$ 
inside  a cone of  opening angle 
$R = \sqrt{\Delta \phi^2 +  \Delta \eta^2 }$ relative to the 
$p+p$ case. 
Simulations of this effect in Fig.~\ref{fig9-10}b at LHC 
bulk matter densities show that this effect is very 
small $\sim 5 \%$  even for narrow cones $R=0.3$~\cite{Salgado:2003rv}.
Perturbative calculations of di-hadron fragmentation also 
find insignificant alteration of the near-side correlation 
function~\cite{Salgado:2003rv}.

\begin{figure}[t!]
\begin{center}
\includegraphics[width=1.8in,height=1.7in]{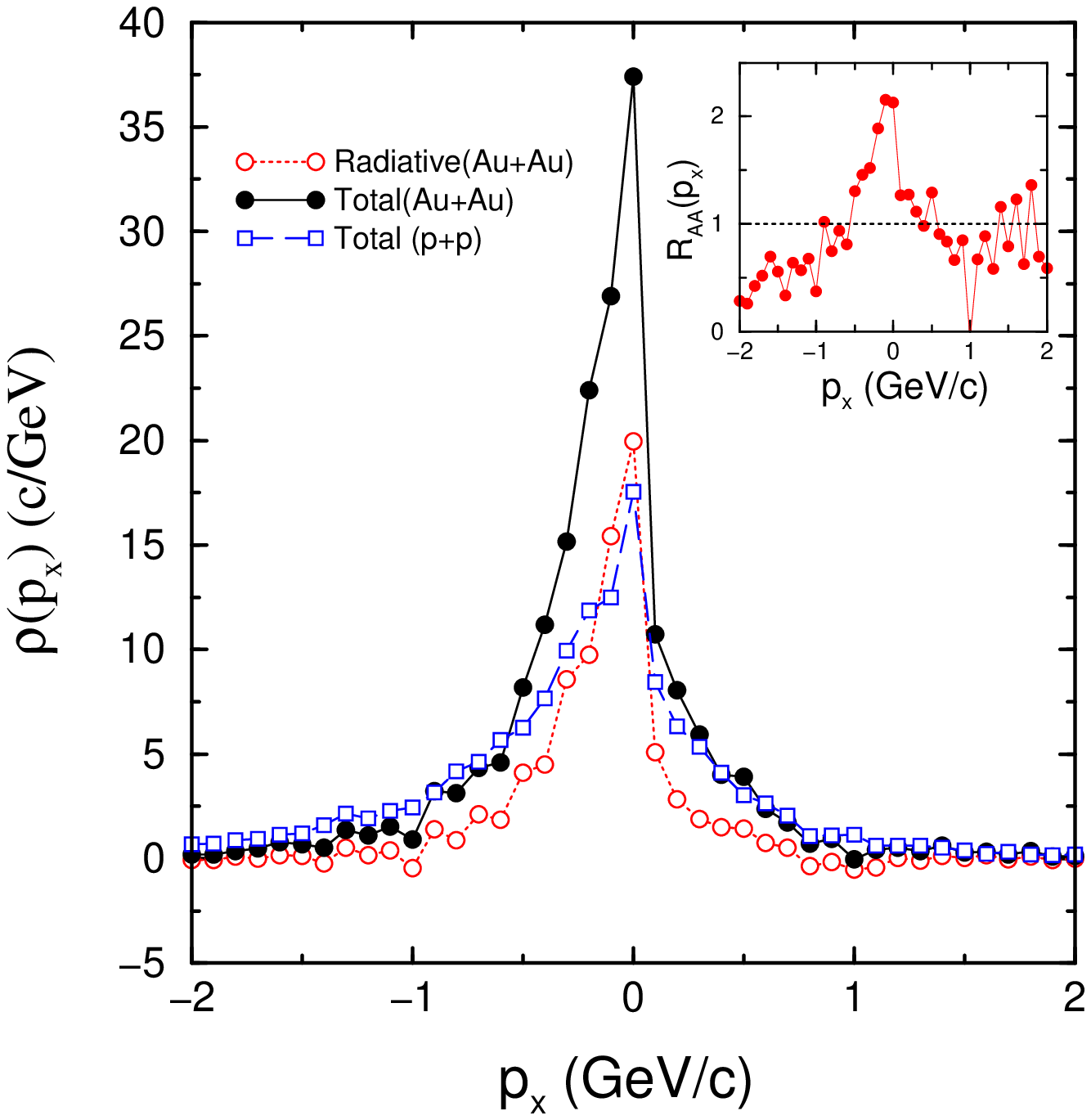}
\hspace*{0.9cm}\includegraphics[width=2.in,height=1.6in]{VitevReappear.eps} 
\end{center}
\vspace*{-0.3cm}
\caption{ (a) From~\cite{Pal:2003zf}: momentum density 
of hadrons associated with  energetic back-to-back jets with and 
without medium induced bremsstrahlung.
(b) From~\cite{Pal:2003zf}: induced partonic multiplicity 
as a function of the 
experimental $p_{T\, {\rm cut}}$ for energetic quark jets.} 
\label{fig11-12}
\end{figure}

While the approximate collinearity of the medium induced gluon 
bremsstrahlung suggests that modifications of the jet shape
may be difficult to measure experimentally, a more promising way 
to verify the predictions of the non-Abelian energy loss 
theory~\cite{Gyulassy:2003mc} may be to search for the ``remnants of 
lost jets''~\cite{Pal:2003zf}. The energy radiated off the leading 
parton is redistributed in the system, 
possibly within the jet cone itself.  If it is subsequently fully 
thermalized, the increase in the soft gluon multiplicity can be estimated
as  $ \Delta N^h \approx \frac{1}{4} \Delta S = \frac{1}{4} 
\frac{\Delta E}{T}$. 
Figure~\ref{fig11-12}a shows the momentum density of hadrons 
$\rho(p_x)$ associated with back-to-back jets   
embedded in a parton cascade~\cite{Pal:2003zf} and the reappearance
of the lost energy at $p_x \approx p_T = 600$~MeV.

\vspace*{0.4cm}

The long  formation times $\tau_f$, which are the basis of the 
non-Abelian LPM effect, may limit the possibility for secondary 
interactions of the radiative quanta. In this case one expects 
fewer harder gluons with $ \omega > 
\omega_{\rm pl}$~\cite{Gyulassy:2000er,Djordjevic:2003zk,Gyulassy:2001nm}.   
As a function of the experimental $p_{T \rm cut}$ the accessible 
$N_{\rm parton}(p_{T \rm cut} ) = 1|_{E-\Delta E \geq p_{T \rm cut} }
 +  P_0(\bar{N}_{\rm g})^{-1}
 \sum_n  n  P_n(\bar{N}_{\rm g}) |_{\Delta E/n \geq p_{T \rm cut}} $, 
with  $\bar{N}_{\rm g}$ and the  probability 
distribution $P_n(\bar{N}_g)$  computed numerically as 
in~\cite{Gyulassy:2001nm,Vitev:2002pf}. 
Figure~\ref{fig11-12}b shows that already at $p_T \simeq 1.5$~GeV a 
large fraction of the induced multiplicity is recovered
and for $\lim_{p_T {\rm cut } \rightarrow 0} E(p_T)  = E^{\rm tot}_{\rm jet}$
and  $\lim_{p_T {\rm cut }\rightarrow 0}  N_{\rm parton}(p_{T \rm cut})
=  \bar{N}_{\rm g}(E_{\rm jet}) +1 $.
I propose that upcoming comparison of the low-$p_T$ experimental 
data on hadron multiplicities associated with energetic di-jets  
to theoretical  calculations may provide an experimental handle on 
the degree of thermalization of the medium induced semi-hard 
bremsstrahlung.

{\bf Summary and conclusions.}
In this talk I emphasized the quantitative applications 
of the theory and phenomenology of medium-induced non-Abelian 
energy loss in reactions involving nuclei.

In QCD processes cold nuclear matter can be
systematically characterized by its scattering 
and absorptive power. Table~\ref{table-cold}, adapted 
from~\cite{Accardi:2003gp}, collects results from 
recent studies of semi-inclusive DIS hadron attenuation, 
Drell-Yan broadening and the Cronin enhancement.   
Within the uncertainties inherent in the theory of strong 
interactions  different approaches consistently find   
$ \frac{\mu^2}{\lambda_g} = 0.1-0.14  \; 
\frac{\rm GeV^2}{\rm fm} $ and final state  
$\langle -\frac{dE_q}{dz}\rangle \; 
= 0.4 - 0.6 \;\frac{\rm GeV}{\rm fm}$ for large nuclei.

In ultrarelativistic heavy ion reactions,  results of the  
jet tomographic analysis~\cite{Vitev:2002pf}  based on the 
assumptions of local thermal equilibrium and longitudinal 
Bjorken expansion are presented in Table~\ref{table-hot}.
For their interpretation,  we recall that lattice 
calculations with improved Wilson and 
staggered fermions suggest $T_c = 171 \pm 4$~MeV for 2-flavor QCD, 
$T_c = 154 \pm 6$~MeV for 3-flavor QCD and 
$\epsilon_c = (0.5-1)\;\frac{\rm GeV}{\rm fm^3}$~\cite{Karsch:2001vs}.
Although the energy density at SPS may have reached or even exceeded 
the critical value for a  phase transition, the 
lifetime of the QGP  $\tau_{\rm tot} - \tau_0 \leq 1 \; {\rm fm}$
seems insufficient to clearly develop many of the signatures
of deconfinement and collectivity. In contrast, at RHIC 
$T_0 = 360 - 410 \; {\rm MeV} \geq 2 T_c$, $\epsilon_0 = 12-20 
\; \frac{\rm GeV}{\rm fm^3} \sim 100 
\, \epsilon_{\rm cold}$ and $\tau_{\rm tot} 
- \tau_0 \sim 5-7 \; {\rm fm}$
is comparable to the transverse extent of the system itself.
These initial conditions are in accord with the 
hydrodynamic estimates~\cite{Kolb:2003dz} based on the 
bulk properties of the plasma.    
Even a sizable theoretical error in the determination of 
$\frac{dN^g}{dy}$ is unlikely to qualitatively 
change RHIC's status on the QCD phase diagram.  
One should also note the much higher $T_0$, $\epsilon_0$ 
and the  significantly longer lifetime of the QGP phase
expected at the LHC.

In conclusion, jet tomographic analysis at RHIC presents 
strong evidence for the creation of the deconfined state of QCD. 
Future jet cone, di-jet correlation and energy balance studies 
are expected to map out the properties of the plasma and reveal 
details of the QCD multiparton dynamics in ultradense nuclear 
matter.

{\bf Acknowledgments.} This  work  is  supported by  the  
United  States  Department of Energy  under Grant   
No. DE-FG02-87ER40371.

\begin{table}[!t]
\caption{\label{table-cold} 
Summary of the transport and absorptive properties of cold nuclear 
matter of energy density
$\epsilon_{\rm cold} = m_N / (\frac{4}{3} 
\pi r_0^3) = 0.14\; \frac{\rm GeV}{{\rm fm^3}} $
adapted from Ref.~\cite{Accardi:2003gp}.}
\begin{indented}
\item[]\begin{tabular}{@{}lll}
\br
&   $ \frac{\mu^2}{\lambda_g}  \; \left[\frac{\rm GeV^2}{\rm fm}\right]$ 
&   $\langle -\frac{dE_q}{dz}\rangle \; \left[\frac{\rm GeV}{\rm fm}\right]$ \\
\mr 
Semi-inclusive DIS quenching \ \ \  & \ \  0.12 \ \  & \  \  0.5  \\
Drell-Yan analysis \ \ \  &  \ \  0.14  $\pm$ 0.1 \ \  
&    \ \ 0.6 $\pm$ 0.45  \\
Cronin effect analysis \ \ \  &  \ \  0.10 -- 0.14  \ \  
&   \ \ 0.4 -- 0.6   \\
Theoretical estimates  \ \ \ &  \ \  0.05  \ \ &  \ \ 0.2    \\
\br
\end{tabular}
\end{indented}
\end{table}

\vspace*{-0.3cm}

\begin{table}[!t]
\caption{\label{table-hot} Summary of the properties of hot nuclear matter
extracted via the jet tomographic analysis of Ref.~\cite{Vitev:2002pf}.
To highlight the difference relative to the case of cold nuclear matter
$\langle -\frac{dE_q}{dz}\rangle$ is numerically estimated for a static 
plasma of $\epsilon_0$ for $E_{\rm jet}= 15$~GeV.
}
\begin{tabular}{@{}lllllll}
\br
 &  \   $\tau_0\; [{\rm fm}]$    &  \  $\tau_{\rm tot}\; [{\rm fm}]$   
&   \ $T_0\; [{\rm MeV}]$    
&  \  $\epsilon_0 \; \left[\frac{{\rm GeV}}{\rm fm^3}\right] $      
&  \  $\frac{dN^g}{dy}$   
&  \  $\langle -\frac{dE_q}{dz}\rangle \; 
\left[\frac{\rm GeV}{\rm fm}\right]$    \\
\mr
SPS \ \ \  &  0.8  &  1.3 -- 2.3 & 205 -- 245  &  1.2 -- 2.6 
& 200 -- 350 &  0.75 -- 1.3  \\
RHIC &  0.6  &  5.5 -- 8 &  360 -- 410   &  12 -- 20  
& 800 -- 1200 & 3.2 -- 4.6  \\
LHC  &  0.2  &  13 -- 23 &  710 -- 850 &  170 -- 350 
& 2000 -- 3500 &  21 -- 35 \\
\br
\end{tabular}
\end{table}


\end{document}